\documentclass[a4paper,fleqn]{cas-sc}

\usepackage[authoryear,longnamesfirst]{natbib}
\usepackage{lineno,hyperref}
\usepackage{amsfonts}
\usepackage{amsmath}
\usepackage{amsthm}
\usepackage{bbm}

\def\tsc#1{\csdef{#1}{\textsc{\lowercase{#1}}\xspace}}
\tsc{WGM}
\tsc{QE}
\tsc{EP}
\tsc{PMS}
\tsc{BEC}
\tsc{DE}
\theoremstyle{plain}
\newtheorem{thm}{Theorem}[subsection]

\theoremstyle{plain}
\newtheorem{thms}{Theorem}[subsection]

\theoremstyle{plain}
\newtheorem{thme}{Theorem}[subsubsection]

\theoremstyle{proposition}
\newtheorem{prop}[thm]{Proposition}

\theoremstyle{assumption}
\newtheorem{assup}[thms]{Assumption}

\theoremstyle{definition}
\newtheorem{defi}[thme]{Definition}

\newproof{pf}{Proof:}
\newproof{pop2}{A.3.2.1 Proof of Proposition \ref{mamam}}
\newproof{pop3}{A.3.3.2.1 Proof of Proposition \ref{maxwel}}
\newproof{pop4}{A.3.3.2.2 Proof of Proposition \ref{papito}}
\newproof{pop5}{A.3.4.1 Proof of Proposition \ref{mamut}}
\newproof{pop6}{A.3.4.2 Proof of Proposition \ref{mamuta}}
\newproof{pop7}{A.3.5.1 Proof of Proposition \ref{greateravg}}
\newproof{pop8}{A.3.5.2 Proof of Proposition \ref{reda}}

\numberwithin{equation}{subsection}

\begin{document}
\let\WriteBookmarks\relax
\def\floatpagepagefraction{1}
\def\textpagefraction{.001}
\shorttitle{Nonlinear Pricing with Misspecified and Arbitrary Perception of the Marginal Price}
\shortauthors{Diego Alejandro Murillo Taborda}

\title [mode = title]{Nonlinear Pricing with Misspecified and Arbitrary Perception of the Marginal Price}                      
\tnotemark[1,2]

\tnotetext[1]{This document is result of my master dissertation at Paris School School of Economics (PSE). My living expenses during this program were financed by the French Government with the Eiffel excellence scholarship.}
\tnotetext[2]{I declare that I don't have relevant or financial interests that relate to the research described in this paper, "Nonlinear pricing with misspecified and arbitrary perception of the marginal price".}
\tnotetext[3]{I would like to thank the Paris School of Economics's staff and teachers for its pedagogical support over my master studies, and particularly to Olivier Compte for his suggestions of literature about behavior economics that could be useful for this project. Also, I want to thank the French goverment for financing my stay in Paris through the Eiffel Excellence Scholarship Program. I'm indebted with professor David Martimort for his patience and invaluable guidance during this project, and I want to thank professor Catherine Bobcheff to be referee of this work. I want to thank Aloisio Araujo (IMPA) for his advice to focus my career on applied mathematics to model and analyze problems in economics. All the mistakes are mine.}

\author[1]{Diego Alejandro Murillo Taborda}[type=editor,
                        auid=000,
                        orcid=0000-0003-4809-1477]
\cormark[1]
\fnmark[1]
\ead{murillo.diego@psestudent.eu}

\address[1]{Paris School of Economics, 48 Boulevard Jourdan, 75014 Paris}

\begin{abstract}
In the context of nonlinear prices, the empirical evidence suggests that the consumers have cognitive biases represented in a limited understanding of nonlinear price structures, and they respond to some alternative perceptions of the marginal prices. In particular, consumers usually make choices based more on the average than the marginal prices, which can result in a suboptimal behavior. Taking the misspecification in the marginal price as exogenous, this document analyzes how is the optimal quadratic price scheme for a monopolist and the optimal quadratic price scheme that maximizes welfare when there is a continuum of representative consumers with an arbitrary perception of the marginal price. Under simple preferences and costs functional forms and very straightforward hypotheses, the results suggest that the bias in the marginal price doesn't affect the maximum welfare attainable with quadratic price schemes, and it has a negligible effect over the efficiency cost caused by the monopolist, so the misspecification in the marginal price is not relevant for welfare increasing policies. However, almost always the misspecification in the marginal price is beneficial for the monopolist. An interesting result is that under these functional forms, more misspecification in the marginal price is beneficial for both the consumers and the monopolist if the level of bias is low, so not always is socially optima to have educated consumers about the real marginal price. Finally, the document shows that the bias in the marginal price has a negative effect on reduction of aggregate consumption using two-tier increasing tariffs, which are commonly used for reduction of aggregate consumption.
\end{abstract}

\begin{highlights}
\item Under quadratic utility functions and arbitrary misspecification in the marginal price, a monopolist with quadratic cost function finds profitable to offer quantity premia if and only if the bias in the marginal price is sufficiently large or the marginal costs increases strong enough in the quantity.
\item The misspecification in the marginal price does not always have a significative effect on the efficiency cost caused by the monopolist.
\item For almost all the levels of misspecification in the marginal price, more misspecification in the marginal price is beneficial for the monopolist. 
\item  More misspecification in the marginal price can be beneficial for both the consumers and the monopolist if the level of bias is low, so not always is socially optimal to have educated consumers about the real marginal prices.
\item The bias in the marginal price has a negative effect on reduction of aggregate consumption using two-tier increasing tariffs.
\end{highlights}


\maketitle

\section{Introduction}

A central assumption in economics is that firms and consumers optimize with marginal prices, which in the context of \cite{ahu79} is related with the concept of substantively rationality. However, the real environment could be complex, that an individual is unable to have in mind the true outcome consequent upon each action, and in this later case, \cite{ahu79} says that the individuals are procedural rational if they uses a reasonable rule to choose their actions. In the context of nonlinear prices, the empirical evidence suggests that the consumers are procedural rational and have cognitive biases represented in a limited understanding of nonlinear price structures, and they respond to some alternative perceptions of the marginal prices. In particular, consumers usually make choices based more on the average than the marginal prices, which can result in a suboptimal behavior.
\\
\\
Furthemore, many policies (taxation, enviromental regulation, energy conservation, among others) are based on theoretical assumptions about how rational consumers respond to nonlinear pricing. Then, understand how do consumers respond to nonlinear pricing is a crucial question in many areas of economics, because if consumers respond differently to the hypotheses of the existing models, these policies could lead to different and undesired outcomes.  
\\
\\
Then, it is important to understand how would a principal (e.g tax authority, monopoly, regulation authority, among others) must design optimal nonlinear tariffs to got its goals (redistribution of income, profit maximization, reduction of aggregate consumption, among others), given that the consumers have misspecification and can respond to some alternative perception of the marginal prices, instead the marginal prices. In particular, this document will analyze if the presence of misspecified consumers in the perception of the marginal prices create exploitative opportunities to the principal, and if the presence of these consumers affects the effectiveness of different-tier increasing block marginal prices in the reduction of the aggregate consumption, which are frequently used in natural resources conservation. 
\\
\\
In particular, this dissertation is related with the literature about the following topics, treated in \cite{ahu61}:
\begin{itemize}
    \item \textbf{Exploitative contracts:} A contract is exploitative if a more informed principal finds economically central considerations driving it derive from trying to profit from the agent's mistakes or lack of information about the economic environment.
    \item \textbf{Environment design:} How a principal must design incentives and individuals decision making environments to achieve specific goals, given the possibility of misspecification for the agents about the decision environment.  
    \end{itemize}
    ORGANIZATION OF THIS PAPER: The section $2$ presents the literature review about some empirical evidence which suggests that consumers can respond to alternative perceptions of the marginal price under nonlinear price schemes. The section $3$ will present the theoretical results of this document. In particular, the subsection $3.1$ presents the setup of the model and the subsection $3.2$ characterizes the optimal price scheme for a monopolist when there is a representative consumer with arbitrary perception of the marginal price, which is a generalization of the "average-price" bias and is given by a weighted average of the marginal price below the consumption. The subsection $3.3$ characterizes the quadratic price scheme that maximizes welfare for the economic enviroment of the previous subsection. The subsection $3.4$ analizes if the level of misspecification in the marginal price is beneficial for the monopolist or some consumers. Finally, the subsection $3.5$ show how the presence of consumers with misspecification in the marginal price, makes it difficult for two-tier increasing tariffs to reduce the aggregate consumption compared to the case when all the consumers are rational. All the proofs are on the appendix.
\section{Literature review}

Economic theory provides at least three possibilities about the perception of the prices for the consumers under nonlinear pricing schemes. The standard model of nonlinear budget sets predicts that consumers respond to marginal prices, and according with Saez (1999) and Borenstein (2009), cited by \cite{ahu69}, the rational consumers respond to expected marginal prices in presence of uncertainty about future consumption. Alternatively, Liebman and Zeckhauser (2004), cited by \cite{ahu69}, argue that the consumers may use the average price as an approximation of the marginal price if the cognitive cost of understanding complex pricing is substantial. 
\\
\\
Also, there exists some literature that provides empirical evidence which support the idea that average effects are more important to individuals than marginal effects for decision making. For example, \cite{ahu70} designs an experiment to investigate the tax rate used by individuals when making marginal economic decisions, and the results suggest that individuals face difficulties in calculating marginal effects, because there are at least as many individuals who use the average tax rate as it would be the marginal tax rate, as individuals who use the true marginal tax rate. 
\\
\\
For example, \cite{ahu72} conduct laboratory experiments to gather data, which show that consumers prefer to choose simple linear tariffs over more complex but also more advantageous nonlinear tariff structures, and this result can be addressed through the lenses of behavioral economics, putting into discussion the impact of cognitive biases such as decision framing, risk attitude, and focalization effects. The data of these experiments show that consumers constantly stick to the tariff with the most simple structure, who analyze the nonlinear pricing schemes as linear schemes. 
\\
\\
Also, there are empirical evidence based on natural experiments, which show that the individuals don't respond strongly to the marginal prices. For example, \cite{ahu73} exploits the price variation at spatial discontinuities in electricity service areas, using different nonlinear price structures among the households of Orange County, California, to develop three empirical estimates (one for each variable) to quantify the effect of the marginal prices, the expected marginal prices, and the average prices on the decisions of the consumers, and the author finds strong evidence which show that consumers respond to average prices rather than marginal or expected marginal prices. In this same paper, the author proposes a strategy that estimates the perceived price directly, and he finds that the resulting shape of the perceived price is nearly identical to the average price. Finally, this paper shows empirically that if consumers respond to average prices, nonlinear tariffs with increasing and non constant marginal price may result in a slight increase in aggregated consumption compared with an alternative constant marginal rate, which contradicts the claim that higher marginal prices for excessive consumption can create an incentive for conservation. 
\\
\\
In this same direction, \cite{ahu69} exploits a natural experiment carried out in 2008 in British Columbia, where the electricity tariffs were changed from a flat marginal rate to a two-tier increasing block tariff for some but not all the cities. Going beyond whether consumer respond to marginal or average prices, this paper explores a novel and more fundamental question: to what extend do consumers misunderstand nonlinear prices? The paper finds that almost all the households (85\%) respond to the average price, a small share (7\%) respond to the marginal price, and a small but important share of households (8\%)  misperceived the nonlinear price scheme, in the sense that these later households perceived mistakenly jumps in marginal prices to apply to all consumption, not just incremental in the quantity of threshold for the two-tier increasing tariff. 
\\
\\
Finally, \cite{ahu62} characterize the monopolist's optimal nonlinear pricing scheme given that consumers can respond to average prices instead of the marginal prices. For the consumers with "average-price" bias, the authors consider a rule to update their consumption level according with the order relation between the marginal utility and their perception of the marginal price (in this case, the average price), and if the process has a steady stable state, it will be a Berk Nash equilibrium as developed in \cite{ahu75}, in which every agent chooses an optimal strategy given their beliefs, and the agent's beliefs put probability one on the set of subjective distributions over consequences that are closed (in terms of the Kullback-Leibler divergence) to the true distribution. However, this paper doesn't consider the possibility to have consumers with arbitrary perception of the marginal price.

\section{Model}

This section will develop some theoretical extensions to \cite{ahu62}, according with some empirical facts about how consumers respond to nonlinear pricing schemes, allowing the possibility of an arbitrary perception of the marginal prices, instead of consider rational or "average-price" biased consumers. In particular, it will analyze if a monopolist can exploit the missperception in the marginal prices to increase its profits, and if this missperception offers new highlights in the optimal design of nonlinear pricing schemes to reduce aggregate consumption (which can be important for the natural resources conservation).

\subsection{Model setup}
Suppose that there is a continuum of consumers with valuation type on $\Theta:=[\theta_0,\theta_1] \subseteq \mathbb{R}_+$, which is private knowledge of the consumers, but the principal (monopolist) only knows that it is distributed according to a twice diffentiable log-concave non-atomic cumulative distribution, $F$, with density function $f$. The type-$\theta$ consumer's payoff is 
$$u(q,\theta)-P(q)$$
where $q \geq 0$ is the quantity consumed, and $P(q)$ is the total payment made to the principal for $q$ units. It is assumed that $u(q,\theta)=q\theta+h(q)$ for a  twice differentiable and strictly concave function $h(\cdot)$,   satisfying $h(0)=0$ \hypertarget{appe30}. Also, the monopolist has a convex twice continuously differentiable cost function, $C$, satisfying $C(0)=0$. 
\\
\\
The objetive of the principal is to determine the pricing scheme, $P: \mathbb{R}_+\rightarrow{}\mathbb{R}$ with $P(0)=0$, that maximizes its expected payoff, given that for every type of consumer, the monopolist is facing a representative consumer with arbitrary perception of the marginal price, as \cite{ahu73}. In this paper, the author consider that for a price scheme, $P$, the perception of the marginal price at consumption $q \geq 0$ is given by a weighted average of the marginal prices until consumption $q$, and in particular, it is equal to the following expression:
\begin{equation}\label{ito}
    \widetilde{P'}(q):=\int_0^q P'(q-\epsilon)dF_q(\epsilon)
\end{equation}
where $F_q$ is the cumulative distribution of a random variable with support contained in $[0,q]$, and we will assume that $F_q$ is public knowledge of the monopolist. For example, if $F_q$ is the cumulative distribution of a uniform random variable on $[0,q]$, the perception of the marginal price for the representative consumer is the average price, and if $F_q$ is the cumulative distribution of $\delta_0$, the perception of the representative consumer is the marginal price. If there exists some $\lambda \in  [0,1]$, such that $F_q$ is the cumulative distribution of the random variable $\lambda U_q+(1-\lambda)\delta_0$, where $U_q \sim U([0,q])$, then the perception of the representative consumer is the average price with probability $\lambda$ and it is the marginal price with probability $1-\lambda$. However, the perception of marginal price at \eqref{ito} ecompasses more general arbitrary perceptions of the marginal price, than whose given by a convex combination of the marginal price and the average price.
\\
\\
According with \cite{ahu73}, there are two ways to recover $\widetilde{P'}(\cdot)$ empirically. The first approach is to assume a certain shape of $F_q$ based on economic theory and test if it is consistent with data. The second approach is directly estimate $F_q$ to find $\widetilde{P'}$. However, this is not a problem of this document, and we will assume that $F_q$ is exogenous. 
\\
\\
For a price scheme $P$, every consumer updates its consumption according with the order relation between his marginal utility and its perception of the marginal price. In particular, when the $\theta$-representative consumer evaluates every possible consumption $q>0$, he finds profitable to choose a strictly higher consumption if $u_q(q,\theta)> \widetilde{P'}(q)$, and he finds profitable to choose a strictly lower consumption if $u_q(q,\theta)< \widetilde{P'}(q)$. Starting from $q_0>0$, under mild assumptions, this rule to choose the optimal consumption will converge to some optimal consumption for the $\theta$-representative consumer, $q_\theta^*$, which is a stable steady state of the following dynamical process:
\begin{equation}
    \frac{dq_t}{dt}:=\lambda_{q_t} \left(u_q(q_t,\theta)- \widetilde{P'}(q_t)\right)
\end{equation}
for some $\lambda_{q_t}>0$. As \cite{ahu62} mention, this rule to choose the optimal consumption for every consumer, is an example of a Berk-Nash equilibrium as developed in \cite{ahu75}. In a Berk-Nash equilibrium, the agent chooses an optimal strategy ($q_t$) given their beliefs (the perception of the marginal price in this case), and the agent's beliefs put probability one on the set of subjective distributions over consequences that are closest (in terms of Kullback-Leibler divergence) to the true distribution.   
\\
\\
According with the stable steady states of the rule to choose the optimal consumption for every consumer given its perception of the marginal price, if $ q: \Theta \rightarrow \mathbb{R}_+$ represents the optimal consumption of the representative consumers, the FOC of the $\theta$-representative consumer is:
\begin{equation} \label{ecuac}
\begin{split}
    & q(\theta)[u_q(q(\theta),\theta)-\widetilde{P'}(q(\theta))]=0
\end{split}
\end{equation}
Also, the monopolist has expected profits equal to:
\begin{equation} \label{exp}
    \pi(P):=\mathbb{E}_\theta \left[ (P(q(\theta))-C(q(\theta)))\right]
\end{equation}
To consider the information of the optimal choices of the consumers in the expected profits of the monopolist, we will use an intuitive and elegant method presented in \cite{wilson}, which allows to consider the distribution of the optimal consumptions for all the consumers, given the cumulative distribution of the valuation types, $F$. We will assume that $\widetilde{P'}(q(\cdot))-h'(\cdot)$ is increasing (wich will be obtained ex-post for some functional forms). The $\theta$-representative consumer will consume more than $q \geq 0$ if $\theta > \widetilde{P'}(q)-h'(q)$\footnote{Since $u_q(\cdot, \theta)$ is decreasing and we are assuming that $\widetilde{P'}(q(\cdot))-h'(\cdot)$ is increasing, this inequality is equivalent to say that $u_q(\cdot,\theta)$ and $\widetilde{P'}(q(\cdot))$ have not cut at $q$ or for a lower quantity than $q$}. Then $F(\theta_P(q))$ is the cumulative distribution of the quantities consumed by all the consumers, where $\theta_P(q):=\widetilde{P'}(q)-h'(q)$.

\begin{prop}\label{dist} For a continuously differentiable function, $P: \mathbb{R}_+ \rightarrow{}\mathbb{R}$, such that $\theta_P(\cdot)$ is increasing, $F(\theta_P(\cdot)))$ is a cumulative distribution if the following condition occurs:
\begin{enumerate}
    \item $\lim_{q \to \infty} \theta_P(q)=\theta_1$ or there exists $q^*_P\geq 0$ such that $\theta_P(q^*_P)=\theta_1$.
    \\
\\
\end{enumerate}
\end{prop}
If a price scheme, $P$, satisfies the hypothesis and condition of the previous proposition, we will denote $q^*_P:=\infty$ if $\lim_{q \to \infty} \theta_P(q)=\theta_1$. Then, all the consumers will consume at most $q^*_P$.
\\
\\
Also $\lim_{q \to 0} F(\theta_P(\cdot)))=F\left(\widetilde{P'}(0)-h'(0) \right)$ represents the probability to have consumption equal to $0$ varying the valuation types.

\subsection{Optimal pricing with arbitrary perception of the marginal price}
Assuming that $P$ is differentiable, integration by parts can be used to express the expected profits of the monopolist as
\begin{equation}\label{profita}
\begin{split}
      \pi(P)&= \mathbb{E}_\theta[P(q(\theta))-C(q(\theta))]
        \\&=\int_0^{q_P^*} (P(q)-C(q))dF(\theta_P(q))
      \\&=\int_0^{q_P^*} (P'(q)-C'(q))(1-F(\widetilde{P'}(q)-h'(q)))dq\\ &=\int_0^{q_P^*} (P'(q)-C'(q))\left(1-F\left(\int_0^q P'(q-\epsilon)dF_q(\epsilon)-h'(q) \right) \right)dq \\ &=\int_0^{q_P^*} L\left(q,P'(q),\int_0^q P'(q-\epsilon)dF_q(\epsilon)\right)dq
\end{split}
\end{equation}
where $L\left(q,P'(q),\int_0^q P'(q-\epsilon)dF_q(\epsilon)\right):=(P'(q)-C'(q))\left(1-F\left(\int_0^q P'(q-\epsilon)dF_q(\epsilon)-h'(q) \right) \right)$. The goal of the monopolist is to choose a function $P: \mathbb{R}_+ \rightarrow{}\mathbb{R}$ which maximizes $\pi(\cdot)$. 
\\
\\
Then $\pi(P)$ depends on the distributed delay of the marginal price $\int_0^q P'(q-\epsilon)dF_q(\epsilon)$, so find the optimal price scheme is a mathematical challenge if we don't impose some additional assumptions. There are different methods to characterize the necessary FOC's for the optimal $P$ when the integrand can be expressed as $L(q,P(q),P'(q),P(q-\tau))$ for a fixed $\tau \in \mathbb{R}$ as in \cite{ahu81} and \cite{ahu82}, or the integrand has the functional form $L(q,P(q),P'(q),P(q-\tau), P'(q-\tau))$ for a fixed $\tau \in \mathbb{R}$ as in \cite{ahu80} and \cite{ahu83}. Also \cite{ahu84}, finds necessary FOC's when the integrand is reduced to $L(q,P(q),P'(q),P(q-\tau_1), P'(q-\tau_2))$ for fixed $\tau_1, \tau_2 \in \mathbb{R}$, so the function $L$ can depend on  different delays of $P$ and $P'$. 
\\
\\
In particular, if there exists $\beta \in [0,1]$ such that $\widetilde{P}'(q)=\beta \frac{P(q)}{q}+(1-\beta)P'(q)$ for all $q>0$, the expected profits can be written as:
\begin{equation}
\begin{split}
      \pi(P)&=\int_0^{q_P^*} (P'(q)-C'(q))\left(1-F\left(\beta \frac{P(q)}{q}+(1-\beta )P'(q)-h'(q) \right) \right)dq\\
      &=\int_0^{q_P^*}  H(q,P(q),P'(q))dq
\end{split}
\end{equation}
\\
\\
where $H(q,P(q),P'(q))=(P'(q)-C'(q))\left(1-F\left(\beta \frac{P(q)}{q}+(1-\beta )P'(q)-h'(q) \right) \right)$. If $H \in C^2(\mathbb{R}^3)$, \cite{brunt} argues that if $P: \mathbb{R}_+ \rightarrow \mathbb{R}$ maximizes $\pi(\cdot)$ over all twice continuously differentiable functions satisfying $P(0)=0$, then $P$ satisfies the following FOC's:
\begin{equation}\label{focb}
    \frac{{\partial H(q,P(q),P'(q))}}{{\partial P}}=\frac{{\partial}}{{\partial q}}\frac{{\partial H(q,P(q),P'(q))}}{{\partial P'}} \ \forall q \in  [0,q_P^*]
\end{equation}
\begin{equation}\label{foc2}
    \beta \frac{P(q_P^*)}{q_P^*}+(1-\beta )P'(q_P^*)-h'(q_P^*)=\theta_1
\end{equation}
\begin{equation}\label{foc3}
    \frac{{\partial H(q_P^*,P(q_P^*),P'(q_P^*))}}{{\partial P'}}=0
\end{equation}
Note that the conditions \ref{foc2} and \ref{foc3} imply $P'(q_P^*)=C'(q_P^*)$ if $B \neq 1$.
\\
\\
However, the goal of this document is to characterize the optimal price scheme for an arbitrary perception of the marginal price, and not only for the case in which it is equal to a convex combination of the marginal price and the average price. Furthemore, the optimal quadratic price scheme is not such a bad approximation to the optimal price scheme if the utility and cost functions are well approximated by quadratic functions. Then, the optimal quadratic price scheme under the following straightforward hypothesis will be characterized, which is satisfied by almost all the usual known probability distributions on $[0,q]$\footnote{For example if $F_q$ is the cumulative distribution function of a random variable with arcsine, uniform, Bates, PERT, triangular, or U-quadratic distribution on $[0,q]$, or is the cumulative distribution of a random variable with law $\delta_0$ or $\delta_q$, as well all the convex combinations of the previous distributions.}:
\begin{assup}\label{a3}
Exists $a_1 \geq 0$ such that $\int_0^q \epsilon dF_q(\epsilon)=a_1 q$ for all $q \geq 0$.
\end{assup}
For notation, we will denote with $X_q$ to the random variable determined by $F_q$. Under the previous assumption, an increase in $a_1$ is equivalent to an increase in the expectation of $X_q$ for all $q \geq 0$. Since $\widetilde{P'(q)}=\int_0^q P'(q-\epsilon)dF_q(\epsilon)$, then under the previous assumption, an increase in $a_1$ implies that the representative consumer has a perception of the marginal price that weights more the values quite lower from its current consumption, $q$. Then $a_1$ is a good measure for the size of the bias in the marginal price.
\\
\\
Now, the problem of the monopolist is to choose $A, B \in \mathbb{R}$ such that $P_{A,B}:=\frac{A}{2}q^2+Bq$ maximizes $\pi(\cdot)$ among all the quadratic price schemes. In this case, the profits of the monopolist according with \eqref{profita} are expressed as:
\begin{equation}\label{mami}
\begin{split}
      \pi(P_{A,B})&=\int_0^{q_{P_{A,B}}^*} (Aq+B-C'(q))(1-F(\widetilde{P_{A,B}'}(q)-h'(q)))dq
\end{split}
\end{equation}
where $\widetilde{P_{A,B}'}(q)=\int_0^q [A(q-\epsilon)+B]dF_q(\epsilon)=A(1-a_1)q+B$ and $q_{P_{A,B}}^*=\inf \left\{{q \geq 0: \widetilde{P_{A,B}'}(q)-h'(q)\geq \theta_1}\right\}$. Assuming that the previous infimum is a minimum (and particularly it is a real number), the Leibniz integral rule implies the following\footnote{ $F(\widetilde{P_{A,B}'}(q_{P_{A,B}}^*)-h'(q_{P_{A,B}}^*))=F(\theta_1)=1$}:
\begin{equation}\label{leib1}
    \frac{{\partial \pi(P_{A,B})}}{{\partial A}}=\int_0^{q_{P_{A,B}}^*} q\left [1-F(\widetilde{P_{A,B}'}(q)-h'(q))-(1-a_1)(Aq+B-C'(q))f(\widetilde{P_{A,B}'}(q)-h'(q)) \right]dq
\end{equation}
\begin{equation}\label{leib2}
    \frac{{\partial \pi(P_{A,B})}}{{\partial B}}=\int_0^{q_{P_{A,B}}^*} \left [1-F(\widetilde{P_{A,B}'}(q)-h'(q))-(Aq+B-C'(q))f(\widetilde{P_{A,B}'}(q)-h'(q)) \right]dq
\end{equation}
Then to find the optimal quadratic scheme, the monopolist compares the profits that obtains with $P_{A,B}$ when $A \to \pm \infty$, or $B \to \pm \infty$, or $(A,B)$ is such that $\frac{{\partial \pi(P_{A,B})}}{{\partial A}}=\frac{{\partial \pi(P_{A,B})}}{{\partial B}}=0$.\footnote{The corner solutions are evaluated in addition for those which satisfied the FOC's}
\begin{prop}\label{mamam}
Suppose that the arbitrary perception of the marginal price satisfies assumption \ref{a3}. Also, suppose that the types are uniformly distributed on $\Theta$, and the utility and cost functions are given  by $h(q):=h_1q-\frac{h_2}{2}q^2$, $C(q):=c_1q+\frac{c_2}{2}q^2$ satisfying:
\begin{itemize}
    \item $h_1 \geq 0, c_1\geq 0, c_2\geq 0, h_2 \geq 0$
    \item $c_2+h_2>0$
    \item $\theta_1+h_1-c_1>0$
    \item $a_1<\frac{2}{3}$
\end{itemize}
Then the optimal quadratic price scheme is $P(q)=\frac{(1-a_1)c_2+(3a_1-1)h_2}{2(1-a_1)(2-3a_1)}q^2+\frac{(1-2a_1)(\theta_1+h_1)+(1-a_1)c_1}{2-3a_1}q$, which satisfies \\ $\pi(P)=\frac{(1-a_1)(\theta_1+h_1-c_1)^2q_{P_{A,B}}^*}{6(\theta_1-\theta_0)(2-3a_1)}=\frac{(1-a_1)^2(\theta_1+h_1-c_1)^3}{6(\theta_1-\theta_0)(2-3a_1)[(1-a_1)c_2+h_2]}>0$.
\end{prop}
\begin{pf}
  Appendix \hyperlink{appe4}{A.3.2.1}
\end{pf}
\qed
\\
Then under the assumptions of the previous proposition, the optimal quadratic price scheme exhibits quadratic discounts when $(1-a_1)c_2+(3a_1-1)h_2<0$ and exhibits quadratic premia when $(1-a_1)c_2+(3a_1-1)h_2>0$. Since $(1-a_1)(2-3a_1)$ is decreasing for $a_1<\frac{2}{3}$, the coefficient of the quadratic part of the quadratic optimal price is increasing in $a_1$ if $-c_2+3h_2>0$, so in this case, as the representative consumer has an increase in his bias, the coefficient of the quadratic part in the optimal quadratic price scheme increases and likely it is profitable for the monopolist to offer quantity premia. In any case, it will be quantity premia under the quadratic optimal price scheme if and only if $c_2>\frac{(1-3a_1)h_2}{1-a_1}$, which occurs if and only if the marginal cost increases strong enough (i.e $\frac{c_2}{h_2}$ is strong enough), or the bias in the marginal price price is sufficiently large since $\frac{1-3a_1}{1-a_1}$ is decreasing in $a_1$.

\subsection{Welfare under the optimal price}
\begin{figure}
	\centering
		\includegraphics[scale=.8]{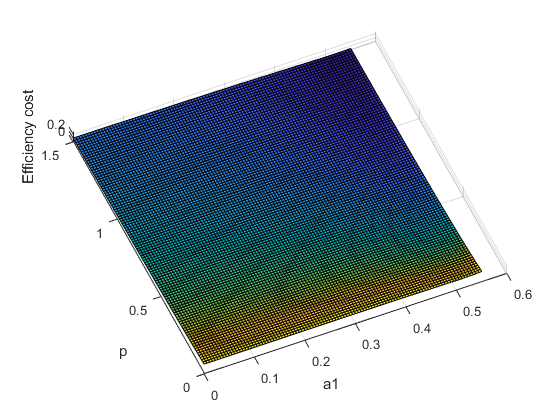}
	\caption{Effect of $p$ and $a_1$ on $\frac{1}{p+1}-F(a_1,p)$}
	\label{FIG:1}
\end{figure}

An interesting question to a regulation authority is to analyze if the optimal price scheme for the monopolist results in a high loss of efficiency, or in a low consumer surplus. This subsection will characterize the efficient pricing schemes when there is a representative consumer with arbitrary perception of the marginal price. Later, this subsection will characterize the efficiency cost of the optimal price schemes for the monopolist, and analyze how the efficiency cost depend on the parameters of the model.
\\
\\
In particular, this subsection will  answer at least the three following questions:
\begin{enumerate}
    \item Is there any relation between the maximum welfare and the maximum profits among all the price schemes?
    \item How big is the efficiency cost when the monopolist is not regulated?
    \item Does the bias in the marginal price have some important effect on the welfare increasing policies?
\end{enumerate}
In this part of the document, it will be used the same economic environment as section $3.2$. For a price scheme $P:\mathbb{R}_+\rightarrow{}\mathbb{R}$ with $P(0)=0$, the total welfares is given by
\begin{equation}
    W(P):=\mathbb{E}_\theta[\theta \Bar{q}(\theta)+h(\Bar{q}(\theta))-C(\Bar{q}(\theta))]
\end{equation}
,where $\Bar{q}(\theta)$ is the optimal consumption of the $\theta$-consumer with arbitrary perception of the marginal price. Using the FOC for the $\theta$-representative consumer \eqref{ecuac} and integration by parts, the total welfare can be expressed as:
\begin{equation}\label{welftot}
\begin{split}
     W(P)&=\mathbb{E}_\theta[[\widetilde{P'}(\Bar{q}(\theta))-h'(\Bar{q}(\theta))] \Bar{q}(\theta)+h(\Bar{q}(\theta))-C(\Bar{q}(\theta))]
     \\
     &=\int_0^{q_P^*} [[\widetilde{P}'(q)-h'(q)]q+h(q)-C(q)]dF(\theta_P(q))
     \\
     &=\int_0^{q_P^*} \left[\left[\frac{{\partial \widetilde{P'}(q)}}{{\partial q}}-h''(q)\right]q+\widetilde{P'}(q)-C'(q)\right][1-F(\widetilde{P}'(q)-h'(q))]dq
\end{split}
\end{equation}
As previously, this subsection characterizes the quadratic price scheme that maximize welfare, $P_{A,B}:=\frac{A}{2}q^2+Bq$, under the very weak assumption \ref{a3}. In this case, the Leibniz integral formula implies the following necessary FOC's\footnote{$\widetilde{P_{A,B}}'(q)=A(1-a_1)q+B$}:
\begin{equation}\label{focaw1}
\begin{split}
     0=\frac{{\partial \mathcal{W}(P_{A,B})}}{{\partial A}}&=\int_0^{q_{P_{A,B}}^*} 2(1-a_1)(1-F(\widetilde{P_{A,B}'}(q)-h'(q)))qdq\\
     &-\int_0^{q_{P_{A,B}}^*} \left[\left[2(1-a_1) A-h''(q)\right]q+B-C'(q)]\right]f(\widetilde{P_{A,B}'}(q)-h'(q)))(1-a_1)qdq
\end{split}
\end{equation}
\begin{equation}\label{focaw2}
\begin{split}
     0=\frac{{\partial \mathcal{W}(P_{A,B})}}{{\partial B}}&=\int_0^{q_{P_{A,B}}^*} (1-F(\widetilde{P_{A,B}'}(q)-h'(q)))dq\\
     &-\int_0^{q_{P_{A,B}}^*} \left[\left[2(1-a_1) A-h''(q)\right]q+B-C'(q)]\right]f(\widetilde{P_{A,B}'}(q)-h'(q)))dq
\end{split}
\end{equation}
\begin{prop}\label{maxwel}
Suppose that the arbitrary perception of the marginal price satisfies assumption \ref{a3}. Also, the types are uniformly distributed on $\Theta$, and the utility and cost functions are given by $h(q):=h_1q-\frac{h_2}{2}q^2$, $C(q):=c_1q+\frac{c_2}{2}q^2$ satisfying:
\begin{itemize}
    \item $h_1 \geq 0, c_1\geq 0, c_2\geq 0, h_2 \geq 0$
    \item $c_2+h_2>0$
    \item $\theta_1+h_1-c_1>0$
    \item $a_1<1$
\end{itemize}
Then the quadratic price scheme that maximizes welfare is $P(q)=\frac{c_2}{2(1-a_1)}q^2+c_1q$, which satisfies $W(P)=\frac{(\theta_1+h_1-c_1)^3}{6(c_2+h_2)}>0$.
\end{prop}
\begin{pf}
Appendix \hyperlink{appe6}{A.3.3.2.1}
\end{pf}
\qed
\\
Then, as usual the price scheme that maximizes welfare depends strongly on the costs. Also, the perception of the marginal price doesn't have effect on the maximum welfare attainable with quadratic price schemes.
\\
\\
The hypotheses of propositions \ref{mamam} and \ref{maxwel} imply a positive relation between the maximum profits and the maximum welfare obtained with quadratic price schemes, although they are not obtained with the same price scheme:
\begin{equation}
    \max_{(A,B) \in \mathbb{R}^2} \pi(P_{A,B})=\frac{(1-a_1)^2(c_2+h_2)}{(2-3a_1)((1-a_1)c_2+h_2)} \max_{(A,B) \in \mathbb{R}^2} W(P_{A,B})
\end{equation}
To characterize the efficiency loss caused by the monopolist, we consider the loss under quadratic price schemes:
\begin{defi}
The efficiency cost caused by a quadratic price scheme, $P$, is defined as $C(P):=\max_{(A,B) \in \mathbb{R}^2} W(P_{A,B})-W(P)\geq 0$. 
\end{defi}
To answer to the questions $2$ and $3$ at the beginning of this section, the following proposition shows that under uniformly distributed types and quadratic utility and cost functions, the bias in the marginal price has an ambiguous effect on the efficiency cost and the welfare under the optimal quadratic price scheme
\begin{prop} \label{papito}
Under uniformly distributed types and the functional forms and hypotheses of propositions \ref{mamam} and \ref{maxwel}, if $W$ represents the welfare under the optimal quadratic price scheme, and we write $c_2=ph_2$ for some $p \geq 0$, then:
\begin{enumerate}
    \item $\frac{{\partial W}}{{\partial a_1}}>0$ if $\frac{1}{3}< a_1<\frac{1}{2}$ and $\frac{{\partial W}}{{\partial a_1}}<0$ if $a_1>\frac{1+\sqrt{73}}{18}$. Since the efficiency cost caused by the monopolist is given by $C:=\frac{(\theta_1+h_1-c_1)^3}{6(\theta_1-\theta_0)(p+1)h_2}-W$, the effect of $a_1$ on $C$ is the negative effect of $a_1$ on $W$.
   \item If $C$ is the efficiency cost caused by the optimal price scheme of the monopolis, then $\frac{{\partial C}}{{\partial p}}<0$ when $a_1 \leq \frac{1}{2}$ and $h_2$ remains constant·
\end{enumerate}
\end{prop}
\begin{pf}
Appendix \hyperlink{appe8}{A.3.3.2.2}
\end{pf}
\qed
\\
 However, under the results of the previous proposition, the effect of $a_1$ on the efficiency cost caused by the monopolist is negligible, and then the perception of the marginal price doesn't have an important effect on the welfare increasing policies, unless the bias in the marginal price is quite high ($a_1 \approx \frac{2}{3}$). On the other hand, a regulation authority finds more attractive to regulated the monopolist when $p$ is near to $0$, so the marginal costs are almost constant, as figure \ref{FIG:1} shows.

\subsection{Exploitative opportunities of the bias for the consumers and the monopolist}
In behavioral economics, there is a growing literature which studies market interaction between sophisticated firms and rational consumers with behaviorally biased consumers. For example, \cite{ahu86} study models where naive consumers overlook add-on prices, or underestimate the chances that they will be subject to hidden fees. Sophisticated consumers exploit the low base prices without purchasing the add-on, and carefully avoid hidden fees. Also, \cite{ahu87}, study a model where consumers may have naive or sophisticated beliefs on their present or future tastes.
\\
\\
In this subsection, it will be analyzed if the level of bias in the marginal price is beneficial for the monopolist and the consumers.The following propositions shows that an increase in the level of bias can be beneficial for the monopolist and the consumers, so not always is socially optimal to have educated consumers about the real marginal price.
\begin{prop}\label{mamut}
For the hypotheses of proposition \ref{mamam}, $a_1$ has a positive effect over the profits under the optimal quadratic price scheme if and only if the optimal quadratic price is convex. In particular, we have the following cases:
\begin{enumerate}
    \item If $c_2 \geq h_2$, then the profits under the optimal quadratic price scheme increase as $a_1$ increases.
    \item If $h_2>c_2$, then the profits under the optimal quadratic price scheme decrease as $a_1$ increases if $a_1<\frac{h_2-c_2}{3h_2-c_2}$, and they increase as $a_1$ increases if $a_1>\frac{h_2-c_2}{3h_2-c_2}$. Then in this case, the minimum profit under the optimal quadratic price schemes is obtained when $a_1=\frac{h_2-c_2}{3h_2-c_2}$.
\end{enumerate}
\end{prop}
\textbf{Proof:} Appendix \hyperlink{appe5}{A.3.4.1.}
\qed
\begin{defi}
For a price scheme $P: \mathbb{R}_+ \rightarrow \mathbb{R}$, the surplus of all the consumers is defined as $CS(P):=W(P)-\pi(P)=\mathbb{E}_\theta[\theta q(\theta)+h(q(\theta))-P(q(\theta))]$, where $q(\theta)$ represents the optimal consumption of the $\theta$-consumer.
\end{defi}
\begin{prop}\label{mamuta}
For the functional forms and hypotheses of proposition \ref{mamam}, under the optimal quadratic price scheme, $P_{A,B}$, if we write $c_2=ph_2$ for some $p\geq 0$, we have the following for the consumer surplus: 
\begin{enumerate}
    \item $\frac{{\partial CS(P_{A,B})}}{{\partial a_1}}>0$ if $p\leq 1.94$ and $a_1<\max \left\{{\frac{1-p}{3-p}, \frac{p-1}{p+1}}\right\}$.
    \item $\frac{{\partial CS(P_{A,B})}}{{\partial a_1}}<0$ if $a_1>0.4$
\end{enumerate}
\end{prop}
\begin{pf}
Appendix \hyperlink{appe11}{A.3.4.2.}
\end{pf}
\qed
\\
Then if $1 \leq p<1.94$, $\max \left\{{\frac{1-p}{3-p}, \frac{p-1}{p+1}}\right\}=\frac{p-1}{p+1}>0$ and $c_2=ph_2 \geq h_2$, so the previous two propositions shows that greater misspecification in the marginal price is both beneficial for the consumers and the monopolist when its level is less than $\frac{p-1}{p+1}$. However, in this case, an increase in the level of the bias increases the expected surplus for all the consumers, but the surplus of the consumers with some valuations could reduce. An interesting exersice is to analyze the distributive effects of an increase in the bias over the consumer surplus of consumers with different valuation type. 
\subsection{Reduction of aggregate consumption with "average-price" bias consumers}
When all the consumers are rational, the theoretical results predict that under the assumption of concave utility functions, the two-tier increasing tariffs\footnote{According with \cite{ahu69}, in a two-tier increasing tariff, consumers pay a low per-unit rate for all consumption up to a defined threshold withing each billing cycle and a higher per-unit rate for all consumption above the threshold} are a effective tool to reduce aggregate consumption, in comparison with a flat marginal price. However, the reduction of the aggregate consumption using two-tier increasing tariffs will be lower as the bias in the marginal price increases.
\\
\\
Suppose that exists $\lambda \in [0,1]$ such that for all $q \geq 0$, $F_q$ in the perception of the marginal price in equation \eqref{ito} is the cumulative distribution of the random variable $\lambda U_q+(1-\lambda)\delta_0$ where $U_q \sim U[0,q]$, so the perception of the marginal price is the average price with probability $\lambda$ and it is the marginal price with probability $1-\lambda$, but unlike the previous sections, this section supposes that the principal can differentiate between rational and "average-price" biased consumers.
\\
\\
Then, the monopolist is facing a mixed population of consumers with a fraction $\lambda \in [0,1]$ of consumers with "average-price" bias, and a fraction $1-\lambda$ of consumers are rational. Using the same idea of subsection $3.1$, presented in \cite{wilson}, assuming that $\frac{P(q)}{q}-h'(q)$ is increasing, the $\theta$-"average-price" biased consumer consumes more than $q> 0$ if $\theta>\frac{P(q)}{q}-h'(q)$. If $P'(\cdot)-h'(\cdot)$ is increasing, the $\theta$-rational consumer consumes more than $q \geq 0$ if $\theta>P'(q)-h'(q)$. Then $F(\Hat{\theta}_P(\cdot))$ and $F(\Tilde{\theta}_P(\cdot))$ are the cumulative distributions of consumption for rational and average-price biased consumers respectively, where $\Hat{\theta}_P(q):=P'(q)-h'(q)$ and $\Tilde{\theta}_P(q):=\frac{P(q)}{q}-h'(q)$. Furthemore, we have the following proposition, which is analogous to proposition \ref{dist}:
\begin{prop}\label{distidu} For a continuously differentiable function, $P: \mathbb{R}_+ \rightarrow{}\mathbb{R}$, such that $\Hat{\theta}_P(\cdot)$ and $\Tilde{\theta}_P(\cdot)$ are increasing, $F(\Hat{\theta}_P(\cdot))$ and $F(\Tilde{\theta}_P(\cdot))$ are cumulative distributions if the two following conditions occur:
\begin{enumerate}
    \item $\lim_{q \to \infty} \Hat{\theta}_P(q)=\theta_1$ or there exists $q^*_P\geq 0$ such that $\Hat{\theta}_P(q^*_P)=\theta_1$.
    \item $\lim_{q \to \infty} \Tilde{\theta}_P(q)=\theta_1$ or there exists $\Tilde{q}^*_P\geq 0$ such that $\Tilde{\theta}_P(\Tilde{q}^*_P)=\theta_1$.
     \\
\\
\end{enumerate}
\end{prop}
If a price scheme, $P$, satisfies the hypothesis and conditions of the previous proposition, we will denote $q^*_P:=\infty$ if $\lim_{q \to \infty} \Hat{\theta}_P(q)=\theta_1$ and $\Tilde{q}^*_P:=\infty$ if $\lim_{q \to \infty} \Tilde{\theta}_P(q)=\theta_1$. Then, all rational consumers will consume at most $q^*_P$, and all "average price" bias consumers will consume at most $\Tilde{q}^*_P$. Then, all type of consumers will consume at most $\max \left\{{q^*_P, \Tilde{q}^*_P}\right\}$.
\\
\\
For a price scheme $P:\mathbb{R}_+\rightarrow{}\mathbb{R}$ satisfying the hypotheses and conditions of proposition \ref{dist}, the aggregate consumption under $P$ will be given by:
\begin{equation}\label{consumo}
    Q(P):=(1-\lambda) \int_0^{q_P^*} qdF(\Hat{\theta}_P(q))+\lambda\int_0^{\Tilde{q}_P^*} qdF(\Tilde{\theta}_P(q))
\end{equation}
We will compare the aggregate consumption among a flat marginal price scheme, $P_1(q):=p_1q$, and a two-tier increasing tariff, $P_2(q):=p_2q\mathbbm{1}_{[0,\Bar{q}]}(q)+[p_2\Bar{q}+p_3(q-\Bar{q})]\mathbbm{1}_{[\Bar{q},\infty]}(q)$, where $\Bar{q}>0$ represents the threshold of the two-tier increasing tariff, and it will be assume that $p_2 \leq p_1 <p_3$, like figure \ref{FIG:2} shows:
\begin{figure}\hypertarget{appe10}
	\centering
		\includegraphics[scale=.6]{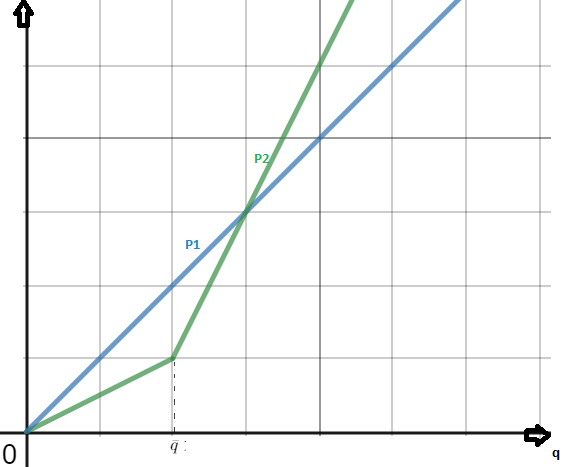}
	\caption{A two-tier increasing tariff and a flat marginal price scheme}
	\label{FIG:2}
\end{figure}
\\
\\
The following proposition shows that the two-tier increasing tariff will result in a strictly higher aggregate consumption if the lowest marginal prices under both price schemes are different and the maximum valuation is inside an intermediate range.
\begin{prop}\label{greateravg}
Consider the model of the subsection $3.1$ with a mixed population of rational consumers and consumers with "average-price" bias, such that the utility function of the $\theta$-type consumer is given by $u(q,\theta):=\theta q+h(q)$ for some differentiable and strictly concave function $h:\mathbb{R}_+\rightarrow{}\mathbb{R}$ satisfying $\lim_{q \to \infty} h'(q)=-\infty$\footnote{This hypothesis is only necessary to satisfy conditions of proposition \ref{distidu}}. If $p_2-h'(\Bar{q})\leq \theta_1 < p_1-h'(\Bar{q})$, then $Q(P_2)>Q(P_1)$.
\end{prop}
\begin{pf}
Appendix \hyperlink{appe13}{A.3.5.1}
\end{pf}
\qed
\\
Furthermore, if the lower marginal price of the two different price schemes is the same (i.e $p_1=p_2$), the two-tier increasing tariff will reduce the aggregate consumption, but the reduction will decrease as the fraction of consumers with "average-price" bias increases as the following proposition shows:
\begin{prop}\label{reda}
Consider the model of the subsection $3.1$ with the utility function of proposition \ref{greateravg}. If $p_1=p_2$, then $Q(P_1)\geq Q(P_2)$, and $Q(P_1)-Q(P_2)$ is decreasing in the fraction of consumers with "average-price" bias.
\end{prop}
\begin{pf}
Appendix \hyperlink{appe13}{A.3.5.1}
\end{pf}
\qed
\\
Then if the fraction of consumers with "average-price" bias is high, to reduce the aggregate consumption in a given quantity, the increase in the marginal price under the two-tier increasing tariffs must be higher in comparison when the situation in which all the consumers are rational.

\appendix
\section{Appendix}
 
\hypertarget{appe4}{}
\begin{pop2}
Under the functional forms of $h, C$, and $F$, according with \eqref{mami}, the profits are given by
\begin{equation}\label{mamita}
\begin{split}
     (\theta_1-\theta_0)\pi(P_{A,B})&=\int_0^{q_{P_{A,B}}^*} ((A-c_2)q+B-c_1)(\theta_1+h_1-B-(h_2+(1-a_1)A)q)dq \\
     &=(B-c_1)(\theta_1+h_1-B)q_{P_{A,B}}^* \\&+\left[ (A-c_2)(\theta_1+h_1-B)-(B-c_1)(A(1-a_1)+h_2))\right]\frac{{q_{P_{A,B}}^*}^2}{2}
     \\ &-(A-c_2)(A(1-a_1)+h_2)\frac{{q_{P_{A,B}}^*}^3}{3}
\end{split}
\end{equation}
Since $a_1<1$, the integrand of the previous integral is a parabola that opens downwards as a function of $A$ (or $B$), so $\limsup_{A \to -\infty} \pi(P_{A,B}) \leq 0$ and $\limsup_{A \to \infty} \pi(P_{A,B}) \leq 0$ for all $B \in \mathbb{R}$, and $\limsup_{B \to -\infty} \pi(P_{A,B}) \leq 0$ and $\limsup_{B \to \infty} \pi(P_{A,B}) \leq 0$ for all $A \in \mathbb{R}$, so the monopolist compare the higher positive profits that he obtains over all pairs $(A,B)$ with $\frac{{\partial \pi(P_{A,B})}}{{\partial A}}=0=\frac{{\partial \pi(P_{A,B})}}{{\partial B}}$ with $q_{P_{A,B}}^*>0$.  Using $q_{P_{A,B}}^*=\frac{\theta_1+h_1-B}{A(1-a_1)+h_2}$, \eqref{leib1} and \eqref{leib2}, the previous FOC's can be written as:
\begin{equation}\label{holi1}
    (\theta_1+h_1-B)[c_2+h_2-a_1A]-2(B-c_1)[A(1-a_1)+h_2]=0
\end{equation}
\begin{equation}\label{holi2}
    (\theta_1+h_1-B)[-(1-a_1)A+h_2+2c_2(1-a_1)]-3(1-a_1)(B-c_1)[A(1-a_1)+h_2]=0
\end{equation}
Combining the previous equations, we obtain: $$0=(\theta_1+h_1-B)[3(1-a_1)(c_2+h_2-a_1A)+2(1-a_1)A-2h_2-4c_2(1-a_1)]$$
To solve the previous equation, we need $B=\theta_1+h_1$ or $A=\frac{(1-a_1)c_2+(3a_1-1)h_2}{(1-a_1)(2-3a_1)}$.
\\
\\
If $B=\theta_1+h_1$, then \eqref{holi1} or \eqref{holi2} imply $A(1-a_1)+h_2=0$ since $\theta_1+h_1-c_1>0$. In this case, equation \eqref{mamita} implies $\pi(P_{A,B})=0$.
\\
\\
If $A=\frac{(1-a_1)c_2+(3a_1-1)h_2}{(1-a_1)(2-3a_1)}$, then \eqref{holi1} or \eqref{holi2} imply $B=\frac{(1-2a_1)(\theta_1+h_1)+(1-a_1)c_1}{2-3a_1}$. In this case, $q_{P_{A,B}}^*=\frac{\theta_1+h_1-B}{(1-a_1)A+h_2}=\frac{(1-a_1)(\theta_1+h_1-c_1)}{(1-a_1)c_2+h_2}>0$. Calculation of equation \eqref{mamita} shows that the profits are $\pi(P_{A,B})=\frac{(1-a_1)(\theta_1+h_1-c_1)^2q_{P_{A,B}}^*}{6(\theta_1-\theta_0)(2-3a_1)}>0$. Note that in this last case, $\widetilde{P_{A,B}'(q)}-h'(q)=((1-a_1)A+h_2)q+B-h_1$ is strictly since $(1-a_1)A+h2=\frac{(1-a_1)c_2+h_2}{2-3a_1}>0$.
\qed
\end{pop2}
\hypertarget{appe6}{}
\begin{pop3}
 For the price scheme, $P_{A,B}$, according with equation \eqref{welftot}, the welfare is given by:
    \begin{equation}\label{welfi}
        \begin{split}
            (\theta_1-\theta_0)W(P_{A,B})&=\int_0^{q_{P_{A,B}}^*} \left[\left[2(1-a_1) A+h_2-c_2\right]q+B-c_1]\right](\theta_1+h_1-h_2q-A(1-a_1)q-B)dq\\
            &=(B-c_1)(\theta_1+h_1-B)q_{P_{A,B}}^*\\
            &+\left[[2(1-a_1)A+h_2-c_2](\theta_1+h_1-B)-(B-c_1)[(1-a_1)A+h_2] \right]\frac{{q_{P_{A,B}}^*}^2}{2}\\
            &-[2(1-a_1)A+h_2-c_2][(1-a_1)A+h_2]\frac{{q_{P_{A,B}}^*}^3}{3}
        \end{split}
    \end{equation}
    In this case, the FOC's \eqref{focaw1} and \eqref{focaw2}, are reduced to:
    \begin{equation}
        \begin{split}
            0&=\int_0^{q_{P_{A,B}}^*} \left[2(\theta_1+h_1-h_2q-A(1-a_1)q-B)-[2(1-a_1)A+h_2-c_2]q+B-c_1]\right]qdq \\
            &=(2(\theta_1+h_1)-3B+c_1)\frac{{q_{P_{A,B}}^*}^2}{2}+[c_2-3h_2-4(1-a_1)A]\frac{{q_{P_{A,B}}^*}^3}{3}
        \end{split}
    \end{equation}
    \begin{equation}
        \begin{split}
            0&=\int_0^{q_{P_{A,B}}^*} \left[\theta_1+h_1-h_2q-A(1-a_1)q-B-[2(1-a_1)A+h_2-c_2]q+B-c_1]\right]dq \\
            &=(\theta_1+h_1-2B+c_1){q_{P_{A,B}}^*}+[c_2-2h_2-3(1-a_1)A]\frac{{q_{P_{A,B}}^*}^2}{2}
        \end{split}
    \end{equation}
    If $q_{P_{A,B}}^*=0$, then $W(P_{A,B})=0$. If $q_{P_{A,B}}^*>0$, then $q_{P_{A,B}}^*=\frac{\theta_1+h_1-B}{A(1-a_1)+h_2}$, and the previous FOC's are reduced to:
    \begin{equation}
    \begin{split}
         0&=3[2(\theta_1+h_1)-3B+c_1][(1-a_1)A+h_2]+2[c_2-3h_2-4(1-a_1)A](\theta_1+h_1-B)\\
         &=2(\theta_1+h_1-B)\left\{3[(1-a_1)A+h_2]+[c_2-3h_2-4(1-a_1)A]\right\}-3(B-c_1)[(1-a_1)A+h_2]
    \end{split}
    \end{equation}
      \begin{equation}
    \begin{split}
         0=&2[\theta_1+h_1-2B+c_1][(1-a_1)A+h_2]+[c_2-2h_2-3(1-a_1)A](\theta_1+h_1-B)\\
         &=(\theta_1+h_1-B)\left\{2[(1-a_1)A+h_2]+[c_2-2h_2-3(1-a_1)A]\right\}-2(B-c_1)[(1-a_1)A+h_2]
    \end{split}
    \end{equation}
Combining the two previous FOC's, we have the following relation:
\begin{equation}
    0=(\theta_1+h_1-B)(6[(1-a_1)A+h_2]+c_2-6h_2-7(1-a_1)A)=(\theta_1+h_1-B)(c_2-(1-a_1)A)
\end{equation}
Then $B=\theta_1+h_1$ or $A=\frac{c_2}{1-a_1}$. If $B=\theta_1+h_1-B$, then $q_{P_{A,B}}^*=0$, and so the total welfare is $0$. If $A=\frac{c_2}{1-a_1}$, from the previous FOC's, we obtain $B=c_1$. In this case, $q_{P_{A,B}}^*=\frac{\theta_1+h_1-c_1}{c_2+h_2}>0$, and $(\theta_1-\theta_0)W(P_{A,B})=\frac{(\theta_1+h_1-c_1)^3}{2(c_2+h_2)}-\frac{(\theta_1+h_1-c_1)^3}{3(c_2+h_2)}=\frac{(\theta_1+h_1-c_1)^3}{6(c_2+h_2)}>0$.
\qed
\end{pop3}
\hypertarget{appe8}{}
\begin{pop4}
For the optimal quadratic price scheme of proposition  \ref{mamam}, we have
\begin{equation}
    B-c_1=\frac{(1-2a_1)(\theta_1+h_1-c_1)}{2-3a_1}
\end{equation}
\begin{equation}
    \theta_1+h_1-B=\frac{(1-a_1)(\theta_1+h_1-c_1)}{2-3a_1}
\end{equation}
\begin{equation}
    2(1-a_1)A+h_2-c_2=2\left[\frac{(1-a_1)p+(3a_1-1)}{2-3a_1} \right]h_2+(1-p)h_2=\left[\frac{a_1(p+3)}{2-3a_1} \right]h_2
\end{equation}
\begin{equation}
    (1-a_1)A+h_2-=\left[\frac{(1-a_1)p+(3a_1-1)}{2-3a_1} \right]h_2+h_2=\left[\frac{(1-a_1)p+1}{2-3a_1} \right]h_2
\end{equation}
\begin{equation}
q_{P_{A,B}}^*=\frac{(1-a_1)(\theta_1+h_1-c_1)}{((1-a_1)p+1)h_2}
\end{equation}
Then, according with equation \eqref{welfi}, the welfare is given by:
\begin{equation}
    \begin{split}
        &\frac{1}{\theta_1-\theta_0}\frac{(1-a_1)^2(\theta_1+h_1-c_1)^3}{6(2-3a_1)^2((1-a_1)p+1)^2h_2}\left[6(1-2a_1)((1-a_1)p+1)+3(a_1(p+3))(1-a_1) \right]\\
        &-\frac{1}{\theta_1-\theta_0}\frac{(1-a_1)^2(\theta_1+h_1-c_1)^3}{6(2-3a_1)^2((1-a_1)p+1)^2h_2}\left[3(1-2a_1)((1-a_1)p+1)+2((a_1(p+3))(1-a_1) \right]\\
        &=\frac{1}{\theta_1-\theta_0}\frac{(\theta_1+h_1-c_1)^3}{6h_2}F(a_1,p)
    \end{split}
\end{equation}
 where $F(a_1,p):=\frac{(1-a_1)^2}{(2-3a_1)^2((1-a_1)p+1)^2}\left[3(1-2a_1)((1-a_1)p+1)+(1-a_1)(a_1(p+3))\right]$.
 \\
 \\
 Also using a computational tool to find partial derivatives\footnote{https://www.derivative-calculator.net/}, $\frac{{\partial F(a_1,p)}}{{\partial a_1}}=\frac{(a_1-1)((4a_1^3-10a_1^2+8a_1-2)p^2+(-9a_1^3+10a_1^2+a_1-2)p+18a_1^3-15a_1^2+3a_1)}{(3a_1-2)^3(pa_1-p-1)^3}$.
  \\
   \\
 For $a_1<\frac{2}{3}$, we have $3a_1-2<0$, $pa_1-p-1<0$ and $a_1-1<0$. On the other hand, the polynomials $4a_1^3-10a_1^2+8a_1-2$, $-9a_1^3+10a_1^2+a_1-2$, and $18a_1^3-15a_1^2+3a_1$ are negative for $\frac{1}{3}<a_1<\frac{1}{2}$, and are positive for $\frac{2}{3}>a_1>\frac{1+\sqrt{73}}{18}$, so $a_1$ has a positive effect on the welfare under the optimal quadratic price scheme when $\frac{1}{3}<a_1<\frac{1}{2}$, and the effect is negative if $a_1>\frac{1+\sqrt{73}}{18}$.
 \\
 \\
 According with proposition \ref{maxwel}, the perception of the marginal price doesn't affect the welfare under the quadratic price scheme that maximizes welfare, which is given by:
 \begin{equation}
     \frac{(\theta_1+h_1-c_1)^3}{6(\theta_1-\theta_0)(p+1)h_2}
 \end{equation}
 Then $a_1$ has a negative effect on the efficiency cost when $\frac{1}{3}<a_1<\frac{1}{2}$, and a positive effect on it when $a_1>\frac{1+\sqrt{73}}{18}$. To show that the efficiency cost is high when the marginal cost is almost constant, using the same computational tool to find partial derivatives, we have that the term $G(a_1,p)= \frac{1}{p+1}-F(a_1,p)$ satisfies:
 \begin{equation}
 \begin{split}
     \frac{{\partial G(a_1,p)}}{{\partial p}}&=-\frac{(4a_1^5-16a_1^4+25a_1^3-19a_1^2+7a_1-1)p^3+(-4a_1^5+2a_1^4+18a_1^3-29a_1^2+16a_1-3)p^2}{(3a_1-2)^2(p+1)^2((a_1-1)p-1)^3}\\
     &-\frac{(7a_1^5-11a_1^4+9a_1^3-13a_1^2+11a_1-3)p+6a_1^5-17a_1^4+12a_1^3-3a_1^2+2a_1-1}{(3a_1-2)^2(p+1)^2((a_1-1)p-1)^3}
 \end{split}
 \end{equation}
 For $a_1\leq \frac{1}{2}$, we have $((a_1-1)p-1)<0$, and
 \begin{equation*}
     4a_1^5-16a_1^4+25a_1^3-19a_1^2+7a_1-1 \leq 0
 \end{equation*}
  \begin{equation*}
     -4a_1^5+2a_1^4+18a_1^3-29a_1^2+16a_1-3 \leq 0
 \end{equation*}
  \begin{equation*}
     7a_1^5-11a_1^4+9a_1^3-13a_1^2+11a_1-3 < 0
 \end{equation*}
  \begin{equation*}
     6a_1^5-17a_1^4+12a_1^3-3a_1^2+2a_1-1 < 0
 \end{equation*}
 Then for the efficiency cost $C=\frac{(\theta_1+h_1-c_1)^3}{6(\theta_1-\theta_0)h_2} G(a_1,p)$, $\frac{{\partial C}}{{\partial p}}<0$ if $a_1\leq \frac{1}{2}$ and $h_2$ remains constant.
 \qed
\end{pop4}
\hypertarget{appe5}{}
\begin{pop5}
According with proposition \ref{mamam}, we have $0<\frac{{\partial \pi^*}}{{\partial a_1}}$ if and only if 
\begin{equation*}
    \begin{split}
      0&<\frac{{\partial (1-a_1)^2}}{{\partial a_1}}(2-3a_1)[(1-a_1)c_2+h_2]-\frac{{\partial [(2-3a_1)[(1-a_1)c_2+h_2]]}}{{\partial a_1}}(1-a_1)^2  \\
      &=-2(1-a_1)(2-3a_1)[(1-a_1)c_2+h_2]-\left\{ [-3[(1-a_1)c_2+h_2]]-c_2(2-3a_1)\right\} (1-a_1)^2 \\
      &=(1-a_1)\left\{c_2[-2(2-3a_1)(1-a_1)+3(1-a_1)^2+(2-3a_1)(1-a_1)]+h_2[-2(2-3a_1)+3(1-a_1)]\right\} \\
      &=(1-a_1)\left\{c_2(1-a_1)(-4+6a_1+3-3a_1+2-3a_1)+h_2(-4+6a_1+3-3a_1)\right\} \\
      &=(1-a_1)[c_2(1-a_1)+h_2(3a_1-1)]
    \end{split}
    \end{equation*}
    Then $a_1$ has a positive effect over the profits under the optimal quadratic price scheme if and only if $c_2(1-a_1)+h_2(3a_1-1)>0$, which means that the optimal quadratic price is convex, according with proposition \ref{mamam}. On the other hand, we have some of the following cases:
    \begin{enumerate}
        \item If $c_2>3h_2$, $c_2(1-a_1)+h_2(3a_1-1)=c_2-h_2+a_1(3h_2-c_2)>c_2-h_2+\frac{2(3h_2-c_2)}{3}=\frac{3h_2+c_2}{3}>0$.
        \item If $3h_2\geq c_2\geq h_2$, $c_2(1-a_1)+h_2(3a_1-1)=c_2-h_2+a_1(3h_2-c_2)\geq c_2-h_2 \geq 0$.
        \item If $h_2 > c_2$, clearly $0<\frac{h_2-c_2}{3h_2-c_2}<\frac{2}{3}$. If $a_1<\frac{h_2-c_2}{3h_2-c_2}$, then
        $$c_2(1-a_1)+h_2(3a_1-1)=c_2-h_2+a_1(3h_2-c_2)<c_2-h_2+h_2-c_2=0$$
        If $a_1>\frac{h_2-c_2}{3h_2-c_2}$, then
        $$c_2(1-a_1)+h_2(3a_1-1)=c_2-h_2+a_1(3h_2-c_2)>c_2-h_2+h_2-c_2=0$$
    \end{enumerate}
    \qed
\end{pop5}
\hypertarget{appe11}{}
\begin{pop6}
Writting $c_2=ph_2$ for some $p \geq 0$, according with equation \eqref{welfi} and propositions \ref{mamam} and \ref{papito}, the consumer surplus under the optimal quadratic price scheme is given by
\begin{equation}
\begin{split}
    CS&=\frac{(\theta_1+h_1-c_1)^3}{6(\theta_1-\theta_0)h_2}F(a_1,p)-\frac{(1-a_1)^2(\theta_1+h_1-c_1)^3}{6(\theta_1-\theta_0)(2-3a_1)((1-a_1)p+1)h_2}\\
    &=\frac{(1-a_1)^2(\theta_1+h_1-c_1)^3\left\{[3(1-2a_1)-(2-3a_1)]((1-a_1)p+1)+a_1(1-a_1)(p+3) \right\}}{6(\theta_1-\theta_0)(2-3a_1)^2((1-a_1)p+1)^2h_2} \\
    &=\frac{(1-a_1)^2(\theta_1+h_1-c_1)^3\left[(1-3a_1)((1-a_1)p+1)+a_1(1-a_1)(p+3) \right]}{6(\theta_1-\theta_0)(2-3a_1)^2((1-a_1)p+1)^2h_2}\\
    &=\frac{(\theta_1+h_1-c_1)^3H(a_1,p)}{6(\theta_1-\theta_0)h_2}
\end{split}
\end{equation}
where $H(a_1,p):=\frac{(1-a_1)^2\left[(1-3a_1)((1-a_1)p+1)+a_1(1-a_1)(p+3) \right]}{(2-3a_1)^2((1-a_1)p+1)^2}$. Using the same computational tool to find partial derivatives\footnote{https://www.derivative-calculator.net/}, we have:
\begin{equation}\label{derivada}
    \frac{{\partial H(a_1,p)}}{{\partial a_1}}=\frac{(a_1-1)\left[(a_1^3-2a_1^2+a_1)p^2+(-5a_1^2+7a_1-2)p+18a_1^3-24a_!^2+12a_1-2 \right]}{(3a_1-2)^3(pa_1-p-1)^3}
\end{equation}
We will use the following auxiliary function for a part of the numerator in the previous derivative:
\begin{equation}
    f(a_1,p):=(a_1^3-2a_1^2+a_1)p^2+(-5a_1^2+7a_1-2)p+18a_1^3-24a_!^2+12a_1-2
\end{equation}
\\
\\
For $0\leq p\leq 1.94$, computations show that
\begin{equation}
\begin{split}
    f\left(\frac{1-p}{3-p},p \right)&=\left(\frac{(1-p)^3}{(3-p)^3}-2\frac{(1-p)^2}{(3-p)^2}+\frac{(1-p)}{(3-p)} \right)p^2+\left(-5\frac{(1-p)^2}{(3-p)^2}+7\frac{(1-p)}{(3-p)}-2 \right)p\\ &+18\frac{(1-p)^3}{(3-p)^3}-24\frac{(1-p)^2}{(3-p)^2}+12\frac{(1-p)}{(3-p)}-2 \\&=\frac{2p(p+3)^2}{(p-3)^3} \leq 0
\end{split}
\end{equation}
\begin{equation}
\begin{split}
    f\left(\frac{p-1}{p+1},p \right)&=\left(\frac{(p-1)^3}{(p+1)^3}-2\frac{(p-1)^2}{(p+1)^2}+\frac{(p-1)}{(p+1)} \right)p^2+\left(-5\frac{(p-1)^2}{(p+1)^2}+7\frac{(p-1)}{(p+1)}-2 \right)p\\ &+18\frac{(p-1)^3}{(p+1)^3}-24\frac{(p-1)^2}{(p+1)^2}+12\frac{(p-1)}{(p+1)}-2 \\&=\frac{2(7p^3-18p^2+23p-28)}{(p+1)^3} < 0
\end{split}
\end{equation}
Also, for $a_1<\frac{2}{3}$, $a_1-1<0$, $3a_1-2<0$, and $pa_1-p-1<0$. 
For $0\leq p<1.94$, we have $0\leq \max \left\{{\frac{1-p}{3-p}, \frac{p-1}{p+1}}\right\}\leq \frac{1}{3}$. Since the functions $a_1^3-2a_1^2+a_1$, $-5a_1^2+7a_1-2$, and $18a_1^3-24a_1^2+12a_1-2$ are increasing in $ [0, \frac{1}{3})$, we have the following for $0 \leq a_1<\max \left\{{\frac{1-p}{3-p}, \frac{p-1}{p+1}}\right\}\leq \frac{1}{3}$:
\begin{equation}
\begin{split}
    f(a_1,p)=&(a_1^3-2a_1^2+a_1)p^2+(-5a_1^2+7a_1-2)p+18a_1^3-24a_!^2+12a_1-2 \\
    &<f\left(\max \left\{{\frac{1-p}{3-p}, \frac{p-1}{p+1}}\right\},p \right) \\
    &\leq \max \left\{{f\left(\frac{1-p}{3-p},p \right),f\left(\frac{p-1}{p+1},p \right)}\right\}\leq 0
\end{split}
\end{equation}
Then, using the expression \eqref{derivada}, we have $\frac{{\partial H(a_1,p)}}{{\partial a_1}}>0$ if $0\leq p<1.94$ and $a_1<\max \left\{{\frac{1-p}{3-p}, \frac{p-1}{p+1}}\right\}$.
\\
\\
For the another part, it is enough to observe that for $0.4<a_1<\frac{2}{3}$, $a_1^3-2a_1^2+a_1>0$, $-5a_1^2+7a_1-2>0$, and $18a_1^3-24a_1^2+12a_1-2>0$, so $\frac{{\partial H(a_1,p)}}{{\partial a_1}}<0$ for $0.4<a_1<\frac{2}{3}$.
    \qed
\end{pop6}
\hypertarget{appe13}{}
\begin{pop7}
Since $\Hat{\theta}_{P_1}(\Bar{q})=p_1-h'(\Bar{q})>\theta_1$, then $q_{P_1}^*<\Bar{q}$. Also, $\Hat{\theta}_{P_2}(\Bar{q})=\Tilde{\theta}_{P_2}(\Bar{q})=p_2-h'(\Bar{q}) \leq \theta_1$, implies $q_{P_2}^* \geq \Bar{q}$, $\Tilde{q}_{P_2}^* \geq \Bar{q}$. As $F$ is a non-atomic distribution, we have the following according with equation \eqref{consumo}:
\begin{equation}
    \begin{split}
        Q(P_2)&=(1-\lambda) \int_0^{q_{P_2}^*} qdF(\Hat{\theta}_{P_2}(q))+\lambda\int_0^{\Tilde{q}_{P_2}^*} qdF(\Tilde{\theta}_{P_2}(q)) \\
        &>(1-\lambda) \int_0^{q_{P_1}^*} qdF(\Hat{\theta}_{P_2}(q))+\lambda\int_0^{\Tilde{q}_{P_1}^*} qdF(\Tilde{\theta}_{P_2}(q)) \\
        &=\int_0^{q_{P_1}^*} qdF(\Hat{\theta}_{P_2}(q)) \\
        &\geq \int_0^{q_{P_1}^*} qdF(\Hat{\theta}_{P_1}(q))\\
        &=Q(P_1)
    \end{split}
\end{equation}
where in the third line, we used that $\Hat{\theta}{P_2}(q)=\Tilde{\theta}_{P_2}(q)$ for all $0\leq q<\Bar{q}$, and in the fourth line, we used the fact that $\Hat{\theta}{P_2}(q)=p_2-h'(q) \leq p_1-h'(q)=\Hat{\theta}_{P_1}(q)$ for all $0 \leq q<\Bar{q}$.
    \qed
\end{pop7}
\hypertarget{appe14}{}
\begin{pop8}
Since $p_1=p_2$, $\frac{P_1(q)}{q}=P_1'(q) \leq \frac{P_2(q)}{q} \leq P_2'(q)$ for all $q>0$, so $\Tilde{\theta}_{P_1}=\Hat{\theta}_{P_1} \leq \Tilde{\theta}_{P_2} \leq \Hat{\theta}_{P_2}$, and then $\Tilde{q}_{P_1}^*=q_{P_1}^* \geq \Tilde{q}_{P_2}^* \geq q_{P_2}^*$, which implies the following according with the equation \eqref{consumo}:
\begin{equation}
    \begin{split}
        Q(P_2)&=(1-\lambda) \int_0^{q_{P_2}^*} qdF(\Hat{\theta}_{P_2}(q))+\lambda\int_0^{\Tilde{q}_{P_2}^*} qdF(\Tilde{\theta}_{P_2}(q)) \\
        &\leq (1-\lambda) \int_0^{q_{P_1}^*} qdF(\Hat{\theta}_{P_2}(q))+\lambda\int_0^{\Tilde{q}_{P_1}^*} qdF(\Tilde{\theta}_{P_2}(q)) \\
        &\leq (1-\lambda) \int_0^{q_{P_1}^*} qdF(\Hat{\theta}_{P_1}(q))+\lambda\int_0^{\Tilde{q}_{P_1}^*} qdF(\Tilde{\theta}_{P_1}(q) \\
        &= \int_0^{q_{P_1}^*} qdF(\Hat{\theta}_{P_1}(q))\\
        &=Q(P_1)
    \end{split}
\end{equation}
Since $q_{P_1}^*=\Tilde{q}_{P_1}^*, q_{P_2}^*, \Tilde{q}_{P_2}^*$  don't depend on $\lambda$, we have:
\begin{equation}
    \frac{{\partial Q(P_2)}}{{\partial \lambda}}=0
\end{equation}
\begin{equation}
    \begin{split}
        \frac{{\partial Q(P_2)}}{{\partial \lambda}}&=\int_0^{\Tilde{q}_{P_2}^*} qdF(\Tilde{\theta}_{P_2}(q))-\int_0^{q_{P_2}^*} qdF(\Hat{\theta}_{P_2}(q))\\
        &\geq \int_0^{q_{P_2}^*} qdF(\Tilde{\theta}_{P_2}(q))-\int_0^{q_{P_2}^*} qdF(\Hat{\theta}_{P_2}(q))\\
        &\geq \int_0^{q_{P_2}^*} qdF(\Hat{\theta}_{P_2}(q))-\int_0^{q_{P_2}^*} qdF(\Hat{\theta}_{P_2}(q))=0
    \end{split}
\end{equation}
    \qed
\end{pop8}
\printcredits

\bibliographystyle{cas-model2-names}

\bibliography{Manuscript_latex}


\end{document}